# Catalyst preparation for CMOS-compatible silicon nanowire synthesis

Vincent T. Renard, Michael Jublot, Patrice Gergaud, Peter Cherns, Denis Rouchon, Amal Chabli & Vincent Jousseaume

*CEA, LETI, MINATEC, F38054 Grenoble, France*

**Metallic contamination was key to the discovery of semiconductor nanowires[1], but today it stands in the way of their adoption by the semiconductor industry. This is because many of the metallic catalysts required for nanowire growth are not compatible with standard CMOS (complementary metal oxide semiconductor) fabrication processes. Nanowire synthesis with those metals which are CMOS compatible, such as aluminium[2] and copper[3-5], necessitate temperatures higher than 450 C, which is the maximum temperature allowed in CMOS processing. Here, we demonstrate that the synthesis temperature of silicon nanowires using copper based catalysts is limited by catalyst preparation. We show that the appropriate catalyst can be produced by chemical means at temperatures as low as 400 C. This is achieved by oxidizing the catalyst precursor, contradicting the accepted wisdom that oxygen prevents metal-catalyzed nanowire growth. By simultaneously solving material compatibility and temperature issues, this catalyst synthesis could represent an important step towards real-world applications of semiconductor nanowires[6-11].**

Gold has been the historic catalyst[1] for silicon nanowire synthesis as it allows excellent yields of good quality nanowires at low temperatures by Chemical Vapour Deposition (CVD). Unfortunately, gold is prohibited in industrial clean rooms since it degrades the electrical properties of semiconductors. It is therefore necessary to find another catalyst. However, compatible metals usually form a liquid alloy with silicon only at elevated temperatures, and as a consequence the standard Vapor-Liquid-Solid[1] (VLS) nanowire growth does not meet CMOS thermal budget (T < 450 °C). Recently the quest for low growth temperatures was fuelled by the discovery of the Vapour-Solid-Solid (VSS) regime where the metal rich catalyst remains in the solid state[12] (For a review see Ref. 13). However, the synthesis which includes catalyst preparation and nanowire growth itself is limited by thermally activated diffusion of silicon in a metal-rich solid particle. Synthesis therefore necessitates high temperature[2-5]. Also, despite evidence for the coexistence of metal-catalyzed and oxide-assisted nanowire growth at high temperatures[14], it is generally believed that the presence of oxygen prevents metal-catalyzed nanowire growth[2]. On the contrary, Figure 1 illustrates that oxidizing copper before CVD nanowire growth can have a dramatically positive effect on the production of nanowires at low temperatures. The initial oxidation state of the 20 nm copper layer was reproducibly controlled *in-situ* by a deoxidation/oxidation step at the same temperature as the following nanowire growth (see Methods). At T = 400°C nanowires are not produced using oxide-free copper while a sufficiently high oxygen pressure (1 Torr) during the oxidation step yields straight nanowires. At still higher oxygen pressures (5 Torr), worm-like structures are obtained (see supplementary section S1). This strong influence allows extremely low nanowire growth temperature (compared to eutectic temperature ∼800°C) and raises question about the role of oxygen in this nanowire growth. In particular, is this regime a new type of metal-oxide assisted growth or does oxidation only assist the VSS growth?

The nanowires are non-epitaxial on the substrate since we used an amorphous metal diffusion barrier (TaN/Ta) on the wafer to meet CMOS standards. The worm-like wires present a lot of structural defects. They are therefore less interesting from the perspective of electrical applications than the straight ones which are single crystals as revealed by High Resolution Transmission Electron Microscope (HRTEM) images (see

Figure 2a and b). In this publication, we will therefore concentrate on the results obtained at the optimal oxygen pressure for producing straight nanowires. These nanowires have a usual morphology with a catalyst at the tip. Their diameter is between 20 and 100 nm. The determination of the crystal structure of the nanowires necessitated careful investigation. Contrary to previous reports[15,16], the combination of Fast Fourrier Transform (FFT) of HRTEM images (Figure 2c) and Raman Scattering was not enough to discriminate between the usual cubic diamond Si I structure and the exotic hexagonal Lonsdaleite crystalline structure[17,18] (Also known as Si IV). This is due to the nanometre-scale thickness of the wires[19,20] (see supplementary Section S2 for more details). X-ray diffraction measurements showed that the nanowires are in fact Silicon I (Supplementary section S2). Energy Dispersive X-ray (EDX) measurements (Figure 2d) revealed that the wires are free of copper in the detection limit of this technique, that the catalyst contains copper and silicon only and is free of oxygen.

Figure 1 illustrates that a certain amount of oxygen during catalyst preparation is necessary for the growth of nanowires, but its role is unclear since oxygen is not found in the catalyst after nanowire growth (Fig 2f). This excludes the possibility of an oxide assisted growth. Oxidation of copper is a long standing subject[21] as metal oxide is usually undesirable for good electrical contacts. Studies at the nanoscale were recently conducted since copper oxide may have interesting applications in photovoltaics and chemistry[22,23]. At low oxygen exposure, a Cuprous oxide ($Cu_2O$) forms until, at higher exposure, a Cupric (CuO) phase nucleates[22] and eventually forms nanowires[24]. Figure 1d, 1e and Supplementary section S1 confirm this scenario. At the optimum oxygen pressure, the seed layer is completely oxidized to $Cu_2O$ (Fig 1e). Interestingly, the seed layer is still not properly dewetted until silane ($SiH_4$) is admitted in the chamber. We therefore propose the following mechanism for the nanowire synthesis in our case. Upon exposure to $SiH_4$, the very reactive Cuprous oxide[23] chemically activates the formation of copper silicide particles following the reaction

$2SiH_4+3Cu_2O \rightarrow 2Cu_3Si+3H_2O+H_2$ ($\Delta G$=- 43.8 kJ as determined with FactSage®
package) until the oxide is entirely reduced. Silicon nanowires then conventionally nucleate from crystalline silicide particles[3-5]. The size of the catalyst particle being larger than ten nanometres, a reduction to 400°C of their melting temperature is not expected[25] and nanowire growth should occur in the VSS regime. A lower limit of 500°C had been previously empirically predicted by Kalache *et al.* for VSS growth of Si nanowires with 3 nm thick copper seed layer based on measurements of the time delay before nanowire growth can occur (incubation time)[5]. They were able to determine an activation energy of *$E_a$=0.98 eV* which permitted to relate incubation time to the formation of $Cu_3Si$ by diffusion of silicon in copper. Indeed, previous work[26] showed that the growth of $Cu_3Si$ by diffusion is thermally activated with the same activation energy as measured by Kalache *et al.*. They predicted infinitely long incubation times, and therefore impossible nanowire growth, below 500°C. It had been established[26] that the growth kinetics of the silicide phase by diffusion follows the law $x^2=k^2t$, where $x$ is the silicide thickness, $t$ is the time and $k^2 \propto \exp\left(\frac{-E_a}{kT}\right)$ is the reaction constant. We therefore expect two orders of magnitude larger incubation times, at a given temperature, than Kalache *et al.* since our Copper layer is about 10 times thicker. Nevertheless, we achieved nanowire growth at 400°C, because the formation of $Cu_3Si$ no longer relies on diffusion but is chemically activated. With consideration of these two independent incubation methods, we conclude that the synthesis temperature of silicon nanowires with Cu is in fact limited (at least down to 400°C) by catalyst preparation rather than by the other limiting steps expected during nanowire growth itself[12] (gas-phase transport of the Si-containing gas to the wire, precursor decomposition, surface/volume diffusion of Si in the silicide or incorporation of Si to the growing wire).

Finally, the importance of oxygen in the chemistry of the obtained nanowires is further evidenced by the substantial morphological evolution of their tips during conservation in ambient atmosphere. TEM images (Figure 3a) reveal that after several hours in atmosphere the catalyst has evolved to a dense-particle array included in an amorphous matrix. In addition the nanowires have a dark plate at the interface between the amorphous region and the silicon (see Arrow in Fig. 3a). A video from High Angle Annular Dark Field tomographic reconstruction is provided as supplementary information for a better understanding of the tip morphology after exposure to oxygen (See also Supplementary Section 3). Elemental mapping (Figure 3b, c and d) from Energy Filtered TEM reveal that the copper rich region is surrounded by a silicon and oxygen rich region and that the plate is copper rich. Electron Energy Loss Spectroscopy experiments show that the Si $L_{2,3}$-edge, acquired from the amorphous region, is typical of $SiO_2$ (See supplementary Section S4). This stack is characteristic of copper-silicide-assisted oxidation of silicon reported in bulk systems[27]. The oxidation proceeds as a cyclic decomposition ($Cu_3Si+O_2 \rightarrow SiO_2+3Cu$) and formation ($3Cu+Si \rightarrow Cu_3Si$) of $Cu_3Si$ at the $SiO_2$/Si interface as long as oxygen is provided at this interface[27] (Figure 3a). Silicon is consumed on one side and $SiO_2$ is rejected on the other side of the interface. Together with EDX measurements, this result confirms indirectly that the nanowires nucleate from $Cu_3Si$. This oxidation could potentially be used to produce $SiO_2$ nanowires. Also, this issue may become important in the perspective of fabricating devices. Indeed, unless a proper encapsulation is used, their properties may be affected by this phenomenon.

In conclusion, chemically assisted incubation of catalysts appears to be a step towards silicon nanowire growth with full CMOS compatibility. The technique could be used to add new functionality (such as nanoelectromechanical systems and sensors) above integrated circuits.

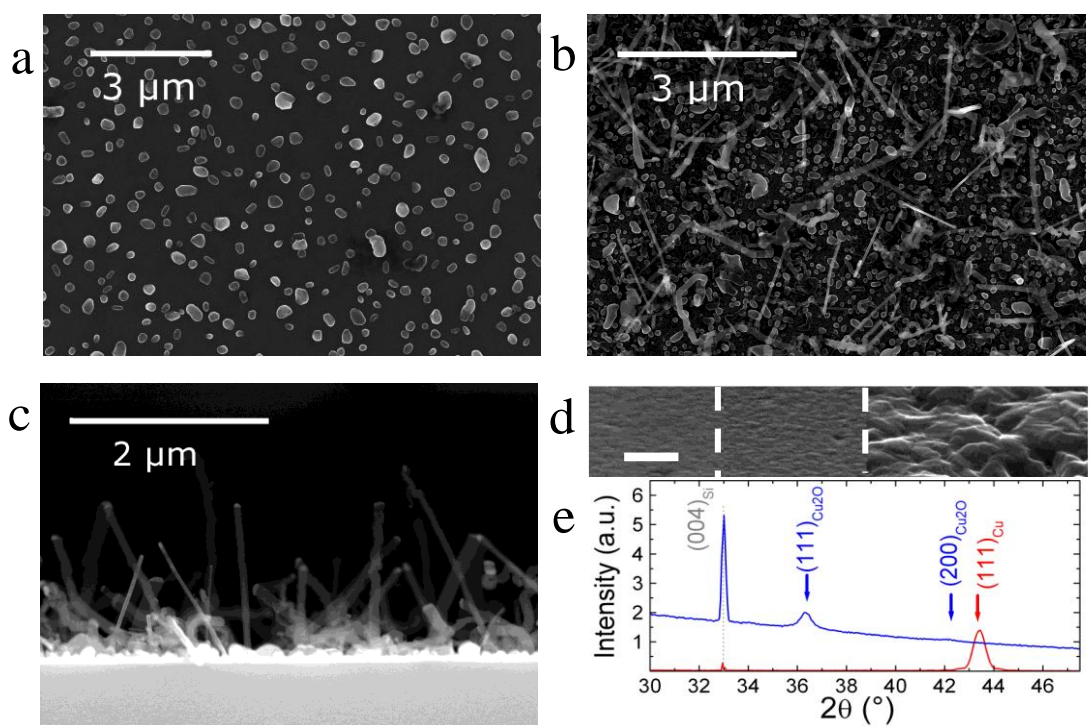

**Figure 1: Silicon nanowire yield at T=400°C with oxide-free and oxidized copper seed layer.**

The growth was performed using 30 Torr of pure $SiH_4$ for 40 minutes. The two examples given differ only by the pressure used during the 180s oxidation of copper after Plasma cleaning (see methods). **a**, Deoxidation only. Oxidation was replaced by 180s under vacuum. **b** and **c**, Oxidation with 1 Torr of oxygen. Top (b) and side (c) view. See Supplementary Section S1 for yields with 5 Torr of Oxygen. **d**, Morphology of the seed layer as deposited (right), after hydrogenated plasma (center) and after oxidation with 1 Torr of oxygen (right). The scale bar is 100 nm. The surface morphology is not affected by plasma cleaning. Oxidation introduces strong roughness but no full dewetting. **e**, XRD spectra of the seed layer before (red) and after (blue) oxidation. It shows that the seed layer is entirely oxidised to $Cu_2O$ since no Cu peak remains after oxidation. The peak noted as $(004)_{Si}$ originates from the silicon substrate.

Page 7 of 20

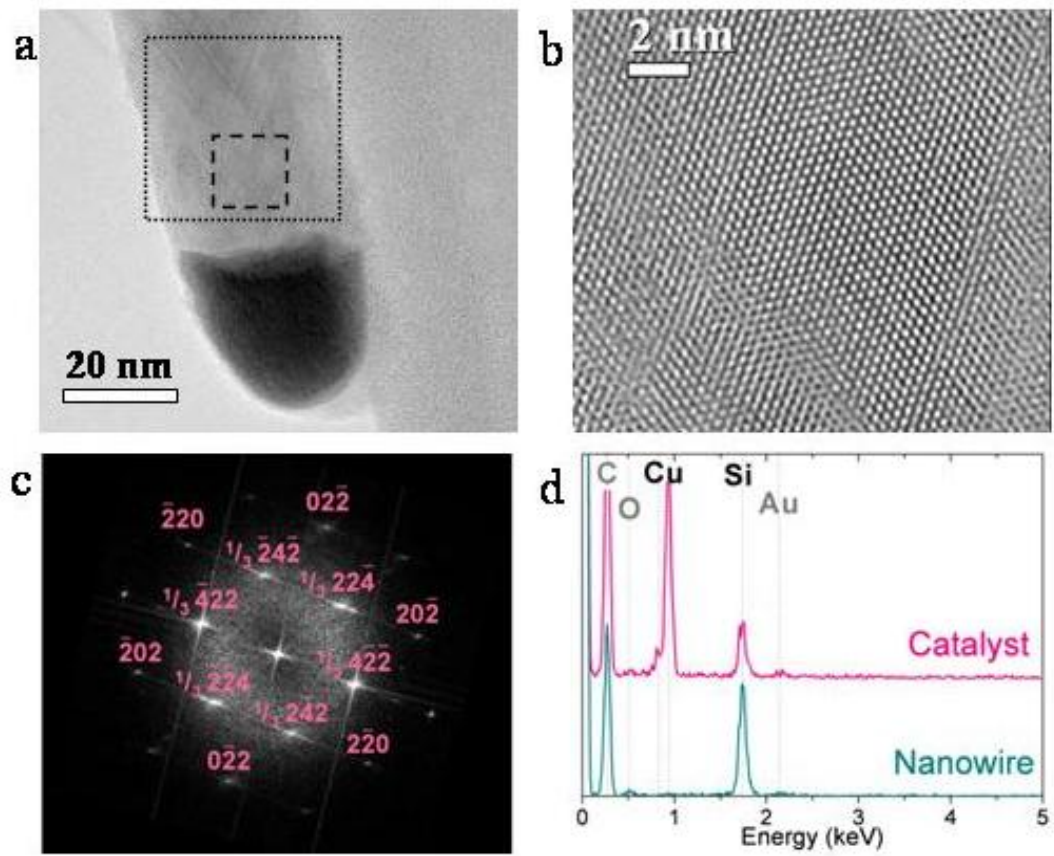

**Figure 2: Structural and chemical characterization of nanowires. a,** HRTEM image of a straight nanowire with the catalyst at the tip. **b,** Magnified HRTEM image (dashed rectangular region in **a**) **c**, FFT performed in the dotted square area shown in **a**. **d**, EDX measurements showing the chemical nature of both the catalyst and the nanowire. The nanowire is free of copper in the detection limit. The catalyst is composed of silicon and copper.

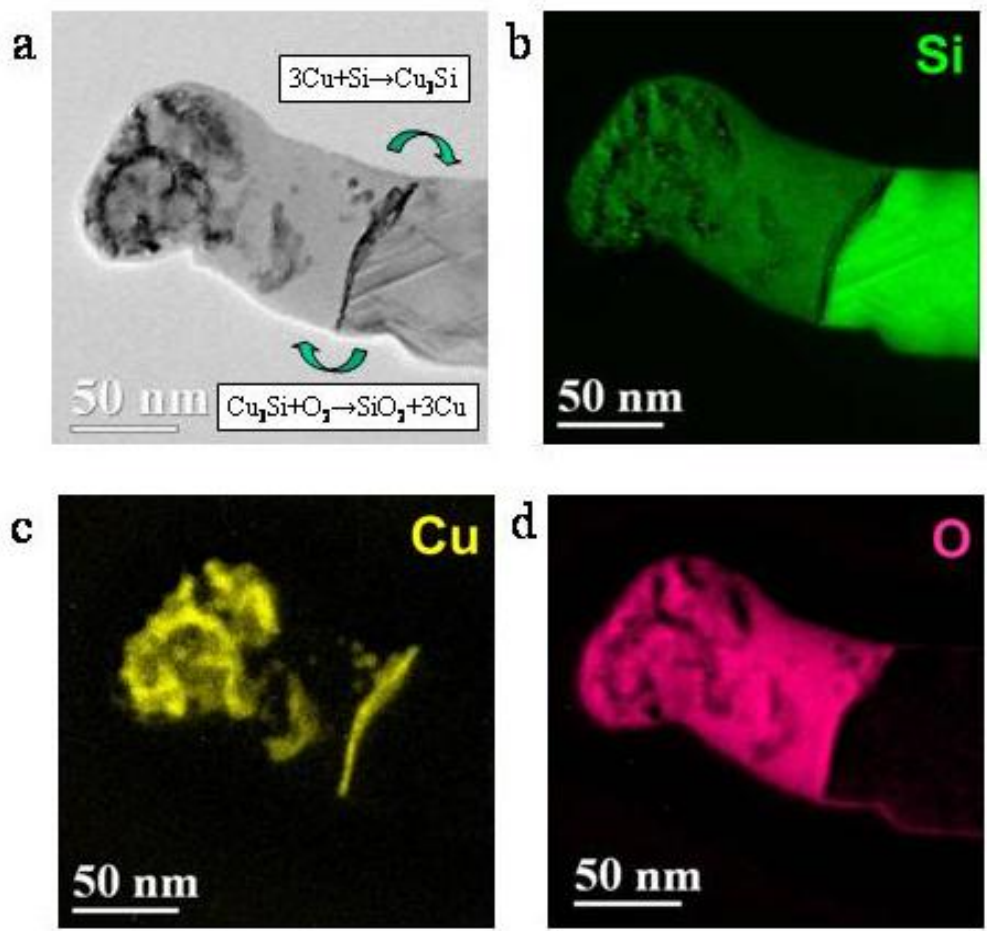

**Figure 3: Oxidation at the tip of the nanowire in ambient atmosphere. a,** TEM image of an oxidized nanowire tip. **b-d,** Location of the different elements illustrated by Energy Filtered EFTEM with bright contrast variation: **b,** Silicon, **c,** copper, **d,** oxygen (see Methods). After a few hours in ambient atmosphere, the copper catalyst region has been transformed in an amorphous matrix containing dense particles. A plate of copper is well distinguished between the $SiO_x$ region and the silicon nanowire. Cyclic reactions described in **a** lead to the displacement of the copper rich interface consuming Si on one side and producing $SiO_2$ on the other side.

## Methods

### Nanowire Growth

Silicon nanowires were prepared by Chemical Vapor Deposition (CVD) in a production 200 mm Applied Materials Centura 5200. Accessible temperatures in the CVD chamber were below 425°C. Standard silicon {100} wafers were used in the experiments. The substrates were first coated with a 10 nm TaN/Ta metal diffusion barrier to avoid metallic diffusion to the substrate (similar results are obtained without this diffusion barrier). A 10 to 50 nm Copper layer was then deposited by a standard PVD technique. The native Copper oxide was first cleaned using hydrogenated plasma[28] in the growth chamber. Copper was then re-oxidized in a controllable way by a flow of pure $O_2$ at pressures between 0.5 and 5 Torr during 180 s. Pure Silane ($SiH_4$) was used as the vapor source of silicon during the growth. Silane pressures were kept between 1 and 50 Torr and growth lasted between one minute and one hour.

### Characterisation

The obtained nanowires were characterized using the Hitachi 5500a scanning electron microscope operating at 30 kV and a JEOL 2010-FEF Transmission Electron Microscope operating at 200kV, equipped with an Omega filter. The nanowires were sonicated in Ethanol and transferred onto carbon grids supported by copper or gold grids for TEM observation. The transfer to TEM was made in less than 20 min to avoid oxidation of the nanowires. Before FFT analysis of the HTREM images, the microscope have been previously calibrated using bulk crystalline Si I samples. The EDX measurements were obtained from nanowires deposited on gold TEM grids. This explains the detection of traces of gold in EDX spectra. The elemental maps were recorded by Energy-Filtered TEM imaging thanks to the omega-filter. Silicon, oxygen and copper maps are obtained by using an energy selecting slit positioned around the energy-losses of the edges : Si $L_{2,3}$ (99 eV), O K (532 eV) and Cu $L_{2,3}$ (931 eV).

X-ray diffraction experiments have been performed on a PANalytical Xpert Pro MRD diffractometer, equipped with a Cu anode in line focus. The upstream optic is a graded W/Si mirror. The equatorial beam divergence is 0.03° and the beam footprint on the sample is defined by a fixed slit of 0.8 mm in the equatorial plane and 10 mm in the vertical plane. The diffracted peaks are registered on a linear detector (PIXcel) used in scanning mode. The instrumental broadening is 0.1°. An offset of 2° is applied in order to decrease the intensity of the 004 substrate reflection. The measurement is performed in $\theta$-$2\theta$ mode.

**Aknowledgements** This work was funded by Carnot UNIFIL and Carnot MULTIFIL projects. We thank D. Lafond and B. Florin for assistance during electron microscopy measurements. We thank C. Charrier and her team for clean room technical assistance. We thank K. Haxaire for diffusion-barrier and metal deposition.

**Author contribution** V.R. and V.J. designed the experiments. V. R. synthesised the nanowires and performed SEM observations. M. J. performed TEM measurements, P.C. performed tomography experiments, D. R. Raman measurements and P. G. XRD measurements. All the authors analysed the data. V. R. and M. J. co-wrote the paper.

# Catalyst preparation for CMOS-compatible silicon nanowire synthesis: Supplementary information

Vincent T. Renard, Michael Jublot, Patrice Gergaud, Peter Cherns, Denis Rouchon, Amal Chabli & Vincent Jousseaume

*CEA, LETI, MINATEC, F38054 Grenoble, France*

**S1 High oxygen pressure during oxidation leads to worm like nanowires**

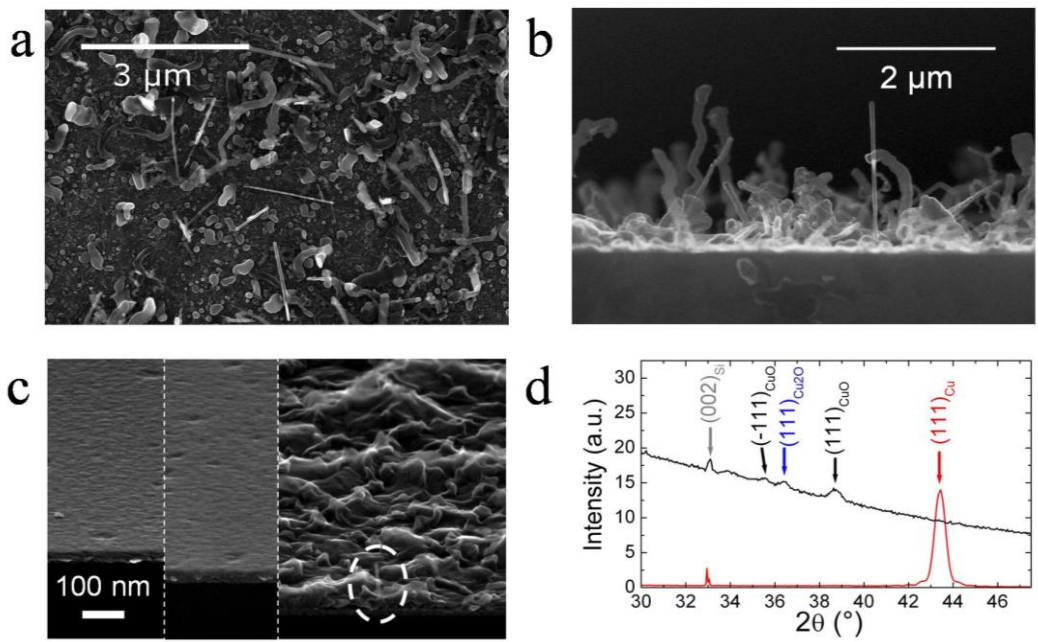

**Supplementary Figure 1:** *a and b, top and side view of the nanowire yield using $P_{O2}$=5 Torr during 180s long oxidation step. c, Surface morphology of the copper seed layer as deposited (left), after plasma cleaning (centre) and after oxidation (right). d, XRD spectra of the seed layer before (red) and after (black) oxidation. The peak marked $(002)_{Si}$ originates from the substrate.*

We used harder oxidation conditions to confirm the crucial role of $Cu_2O$ in the synthesis of silicon nanowires at low temperature. After native oxide removal (see methods), the copper layer was oxidized for 180 s with 5 Torr of pure oxygen. The growth was performed using 30 Torr of pure $SiH_4$ for 40 minutes. Supplementary figures 1a and 1b show SEM micrograph of top view and side view of the yields. The nanowires obtained are worm-like instead of the straight ones seen in Fig. 1b and c. We performed morphological and chemical characterization of the seed to confirm that this change in nanowire morphology is due to the appearance of CuO. The plasma cleaning does not modify the surface morphology of the seed while oxidation strongly reshapes the surface (Suppl. Fig. 1c). XRD measurements (Suppl. Fig. 1d) confirm that Cupric oxide phase appears during oxidation at such high pressure (compared to that used in Fig. 1). This is also illustrated by the nucleation of CuO nanowires (dashed circle in Suppl. Fig. 1c) in agreement with the literature. In turn, the quality of the produced silicon nanowires degrades which offers further evidence that the incubation is controlled by $Cu_2O$.

## S2 Determining the crystal structure of the nanowires

The Fast Fourrier Transform (FFT) of HRTEM image (Suppl. Fig. 2a and 2b) produces a surprising result. Three planes with nearly the same interplanar distance (about 0.32 nm) and the same angle (60°) between them are identified. This does not relate to any known zone axis of the cubic diamond Si I structure (space group Fd3m).

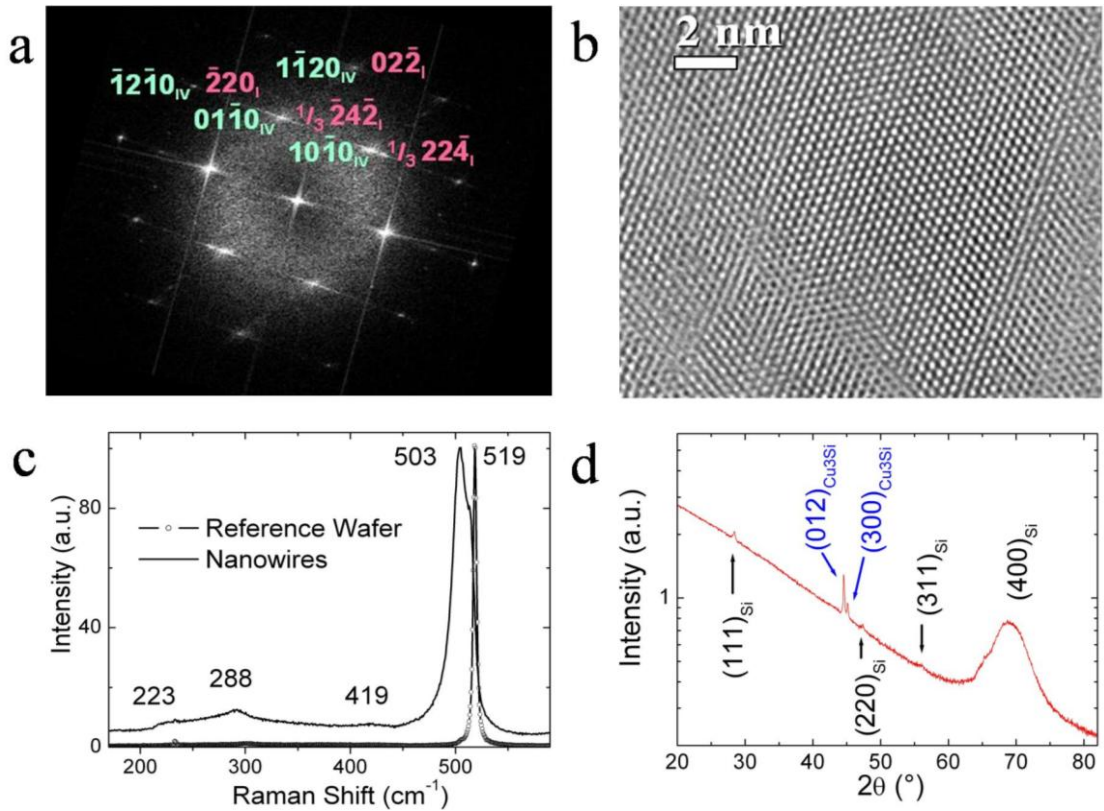

**Supplementary Figure 2:** *a, Fast Fourrier Transform of the high resolution image of a Si nanowire (b) also shown in Fig2b and c. Two possible indexing are represented: Lonsdaleite in blue and cubic diamond in purple (see the text). c, Raman spectrum of the nanowires. The dotted spectrum shows the Raman response of a reference Si wafer with the characteristic peak around 520 cm$^{-1}$. d XRD measurements. Peaks indexed in black correspond to cubic diamond silicon while peaks indexed in blue correspond to Cu$_3$Si. The large intensity of the Cu$_3$Si signal is due to the material at the bottom of the nanowires (See Fig 1b). The broad peak at 70° corresponds to the signal from the substrate.*

The FFT could be explained by classically forbidden 1/3{422} reflections which are present in the [111] zone axis orientation of silicon I for very thin specimens[1] (like nanowires). Alternatively, it could be explained by reflections in the [0001] zone axis of the unusual Lonsdaleite (hexagonal) crystalline structure[2], also known as silicon IV (space group P6$_3$mc). This metastable phase of silicon, discovered[3] in the 1960's, was recently claimed to have been observed in nanowires by VLS with a gold catalyst[4]. Following previous investigation[4,5] we proceeded to use Raman scattering measurements to determine the crystal structure (Suppl Fig 2c) and methods). The resonances at 223 cm$^{-1}$, 288 cm$^{-1}$, 503 cm$^{-1}$ and even 419 cm$^{-1}$ suggest that our nanowires are Si IV[4,5]. However, this set of resonances is also characteristic of 2TA(L), 2TA(X), TO(L) and LO($\Delta$) modes of nanocrystalline Si I[6]. Raman scattering is therefore also unable to distinguish between Si I and Si IV. X-Ray-Diffraction shows that Si I and Cu$_3$Si only are present on the substrate (Suppl Fig 2e). We therefore conclude that our nanowires have the usual cubic diamond structure, though further investigation on a large amount of material will be necessary to eliminate the possibility of a tiny fraction of Si IV nanowires. It follows that it is now possible to correctly index the FFT of the HRTEM image (Fig. 2c). Our observations illustrate that extreme care is needed for nanoscale characterization and raises some concern about previous claims of the observation of Si IV nanowires.

Method

An Ar+ laser with 514.5 nm as the excitations wavelength was used as the photon source during Raman measurements. A 100× objective was used to focus the laser spot and to collect the emitted radiation. A low excitation power (20 kW/cm2) was used to avoid sample heating. The signal was dispersed through a grating spectrometer and projected to a liquid N2-cooled charged coupled device (CCD).

## S3 Nanowire's tip after oxidation: a tomographic reconstruction.

The supplementary video shows an electron tomogram acquired from a nanowire. The video shows first a volume view of the tomogram. A series of vertical slices through the 3D volume are then displayed in sequence. In the isosurface representations, the first threshold used corresponds to the intensity level of the Si, hence showing the outer surface of the lower region of the wire in blue. A second threshold is defined, corresponding to the Cu-rich material in top region of the wire. This is displayed in red, and can be seen when the outer surface is transparent. These representations allow observation of the distribution of the Cu-rich material.

Method:

Electron tomography was carried out on an FEI Titan S/TEM operating at 300kV. A tilt series was acquired consisting of 141 high angle annular dark field (HAADF) images. The sample was tilted between +/- 70°, using a Fischione 2020 high tilt sample holder. The tilt series was reconstructed using the simultaneous iterative reconstruction technique (SIRT) implemented in the FEI Inspect 3D software. Visualisation is carried out using Chimera (software distributed by UCSF).

## S4 Characterizing the amorphous region

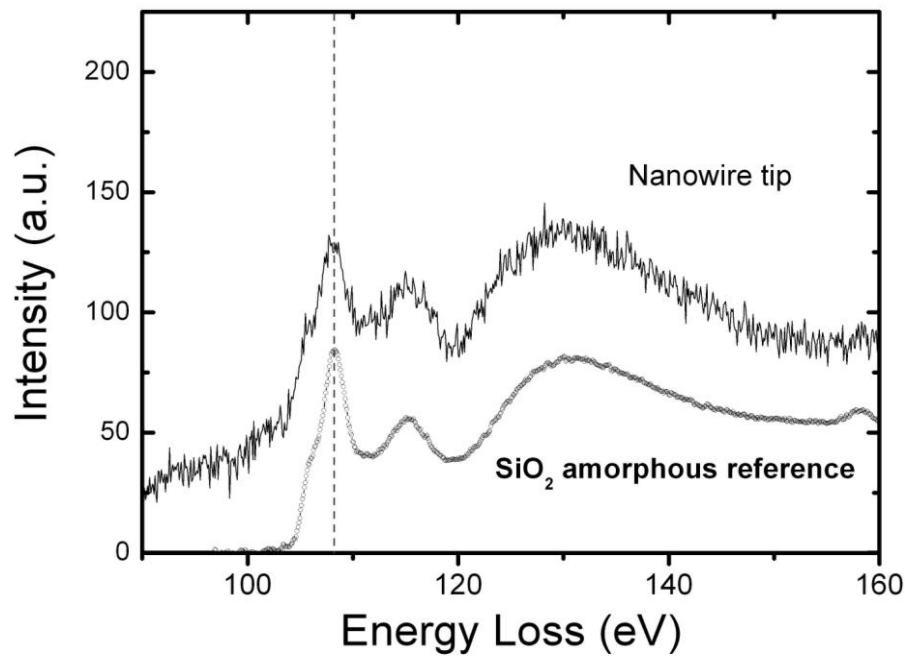

**Supplementary Figure 3** : *EELS fine structure in the amorphous region after tip oxidation. The spectrum at the Si L2,3-edge (solid line) was acquired at the point indicated by an arrow in the inset. It compares very well with the spectrum measured in a reference SiO2 specimen (line with open symbols).*